\documentclass[%
 reprint,
superscriptaddress,
 amsmath,amssymb,
 aps,prd,
]{revtex4-2}

\usepackage{graphicx}% Include figure files
\usepackage{dcolumn}% Align table columns on decimal point
\usepackage{bm}% bold math
\usepackage[mathlines]{lineno}% Enable numbering of text and display math
\usepackage{comment}
\usepackage{hyperref}
\usepackage{cleveref}
\usepackage[usenames]{color}

\newcommand{\jcap}{J. Cosmol. Astropart. Phys.}
\newcommand{\mnras}{Monthly Notices of the Royal Astronomical Society}
\newcommand{\apjl}{the Astrophysical Journal Letters}
\newcommand{\apjs}{the Astrophysical Journal Supplement}

\newcommand{\aap}{Astronomy and Astrophysics}
\newcommand{\pasp}{Publications of the Astronomical Society of the Pacific}

\begin{document}

\preprint{APS/123-QED}

\title{White dwarf binary modulation can help stochastic gravitational wave background search}% Force line breaks with \\

\author{Shijie Lin}
\affiliation{Institute for Frontier in Astronomy and Astrophysics, Beijing Normal University, Beijing, 102206, People's Republic of China}
\affiliation{Department of Astronomy, Beijing Normal University, Beijing 100875, People's Republic of China.}
%%%%%%%%%%%%%%%%%%%%%%%%%%%%%%%%%%
\author{Bin Hu}%
\email{bhu@bnu.edu.cn}
\affiliation{Institute for Frontier in Astronomy and Astrophysics, Beijing Normal University, Beijing, 102206, People's Republic of China}
\affiliation{Department of Astronomy, Beijing Normal University, Beijing 100875, People's Republic of China.}
%%%%%%%%%%%%%%%%%%%%%%%%%%%%%%%%%%
\author{Xue-Hao Zhang}
\affiliation{Lanzhou Center for Theoretical Physics,
	Key Laboratory of Theoretical Physics of Gansu Province,
	School of Physical Science and Technology,
    Lanzhou University, Lanzhou 730000, China} 
\affiliation{Institute of Theoretical Physics \& Research Center of Gravitation, Lanzhou University, Lanzhou 730000, China}
%%%%%%%%%%%%%%%%%%%%%%%%%%%%%%%%%%
\author{Yu-Xiao Liu}
\affiliation{Lanzhou Center for Theoretical Physics,
	Key Laboratory of Theoretical Physics of Gansu Province,
	School of Physical Science and Technology,
    Lanzhou University, Lanzhou 730000, China} 
\affiliation{Institute of Theoretical Physics \& Research Center of Gravitation, Lanzhou University, Lanzhou 730000, China}

\date{\today}

\begin{abstract}
For the stochastic gravitational wave backgrounds (SGWBs) search centred at the milli-Hz band, the galactic foreground produced by white dwarf binaries (WDBs) within the Milky Way contaminates the extra-galactic signal severely. 
Because of the anisotropic distribution pattern of the WDBs and the motion of the spaceborne gravitational wave interferometer constellation, the time-domain data stream will show an annual modulation. This property is fundamentally different from those of the SGWBs.  
In this Letter, we propose a new filtering method for the data vector based on the annual modulation phenomenon. 
We apply the resulted inverse variance filter to the LISA data challenge. 
The result shows that for the weaker SGWB signal, such as energy density $\Omega_{\rm astro}=1\times10^{-12}$, the filtering method can enhance the posterior distribution peak prominently.  
For the stronger signal, such as $\Omega_{\rm astro}=3\times10^{-12}$, the method can improve the Bayesian evidence from `substantial' to `strong' against null hypotheses.  
This method is model-independent and self-contained. It does not ask for other types of information besides the gravitational wave data. 
\end{abstract}

\maketitle

%%%%%%%%%%%%%%%%%%%%%%%%%
\emph{Introduction.--} 
The stochastic gravitational wave background (SGWB) \cite{1999PhRvD..59j2001A,2017LRR....20....2R,2019RPPh...82a6903C} is one of the primary targets of all types of gravitational detectors. There are two kinds of SGWB signals, namely those from astrophysical sources and from cosmological origins. The former is produced by many independent and unresolved continuous gravitational wave sources, such as black hole binaries \cite{2016PhRvL.116m1103A} and neutron star binaries \cite{2018PhRvL.120i1101A}. The incoherent superposition from each of the single sources leads to a stochastic nature of the gravitational wave background \cite{2011RAA....11..369R,2019ApJ...871...97C}. 
The typical energy density of cosmological stochastic gravitational waves is a complex topic that depends on the production mechanisms. 
The energy density of the stochastic gravitational wave background predicted by slow-roll inflation is expected to be very small, $\Omega_{\rm GW}\sim10^{-16}$ with a flat spectrum.
A sizeable cosmological SGWB needs for some exotic mechanisms in the early universe, such as curvature peak, first-order phase transition and cosmic strings \cite{2012JCAP...06..027B,2018CQGra..35p3001C}, {\it etc.} 

The search for SGWB signal is performed in multi-frequency bands. The LIGO-Virgo-KAGRA (LVK) collaboration looks for this signal in 
the frequency range between 10 Hz and 10 $k$Hz, where the optimal sensitivity is centred at 100 Hz. According to the merger rate estimated from binary neutron star GW170817, it is believed that the energy density of SGWB generated from the unresolvable binary neutron stars would be $\Omega_{\rm GW}=1.8^{+2.7}_{-1.3}\times10^{-9}$ at $f_{\rm ref}=25\,\rm Hz$ \cite{2018PhRvL.120i1101A}. Unfortunately, up to the third observing run (O3), there is no detection evidence \cite{2017PhRvL.118l1101A,2019PhRvD.100f1101A,2021PhRvD.104b2004A}. For instance, the upper limit of the dimensionless energy density $\Omega_{\rm GW}\leq 5.8\times10^{-9}$ at the $95\%$ credible level for a flat spectrum in the most sensitive part of the LIGO band ($20-86$ Hz). 
Furthermore, the pulsar timing array (PTA) can be utilised to search SGWB in the nano-Hz band. The recent NANOGrav (the North American Nanohertz Observatory for Gravitational Waves) 12.5-year data reported a strong evidence of a stochastic process. Unfortunately, this process has no statistically significant quadrupolar spatial correlations, which would consider necessary to claim an SGWB detection predicted by the theory of general relativity \cite{2020ApJ...905L..34A}.

The spaceborne GW interferometers like LISA \cite{2017arXiv170200786A}, Taiji \cite{2020IJMPA..3550075R} and TianQin \cite{2016CQGra..33c5010L} are able to detect SGWB centred at the milli-Hz band ($10^{-4}-10^{-1}$Hz). In this frequency range, it is predicted that the SGWB signals from astrophysical sources and cosmological origins have significant evidence to be detected \cite{2018CQGra..35p3001C,2019RPPh...82a6903C,2021MNRAS.500.1421Z}. However,  there is a significant galactic foreground contamination produced by tens of millions of white dwarf binaries (WDBs) in this frequency range\cite{1987ApJ...323..129E,1997CQGra..14.1439B,2001A&A...375..890N,2006PhRvD..73l2001T}. 
In the LISA Data Challenge (LDC) \cite{2008CQGra..25r4026B}, methods to separate the galactic foreground from the SGWB signal are in demand, and some solutions have been proposed \cite{2014PhRvD..89b2001A,2021JCAP...01..059F}. 
To separate and remove the galactic foreground, there are two basic directions \cite{2014PhRvD..89b2001A}. First, the foreground of the Milky Way has different energy spectrum shapes from instrument noise and the typical SGWB models, which can help us effectively distinguish noise and multiple different SGWB components \cite{2021JCAP...05..052P,2021JCAP...01..059F}. It is worth noting that, for the galactic foreground, some high-frequency WDBs can even be identified and eliminated from the data, thus reducing the impact of the galactic foreground at high frequencies \cite{2007PhRvD..75d3008C,2021PhRvD.104b4023Z}. Second, since most of the WDBs in the Milky Way are distributed in the disk \cite{2001A&A...375..890N,2019MNRAS.490.5888L}, the galactic foreground spatial distribution is anisotropic, which is obviously different from the isotropic SGWB. 
The noise intensity generated by the galactic foreground will strongly depend on the direction of the detector constellation. 
Since the angle between the LISA detector plane and the equatorial coordinate system will not change a lot within the total observation time \cite{2017arXiv170200786A}, its time-domain signal will show annual modulation in the one-year observation data \cite{2004PhRvD..69l3005S,2014PhRvD..89b2001A,2021JCAP...01..059F}. 
We can use this property to separate the galactic foreground from the extra-galactic SGWB. Boileau et al. \cite{2021MNRAS.508..803B} estimated the parameters for the different signal classes and measured the orbital modulation of the galactic foreground. However, they did not use such information for separating the SGWB signal.
In this Letter, we utilise the annual modulation to reduce the galactic foreground contamination. The resulting method is nothing but the inverse variance filter (IVF), which is widely used in cosmic microwave background (CMB) data analysis, such as foreground component separation \cite{2007ApJS..170..288H} and gravitational lensing studies \cite{2017PhRvD..96f3510C}. 

%%%%%%%%%%%%%%%%%%%%%%%%%
\emph{Methodology.--} 
For a time-domain data stream, $d(t)$, whose campaign length is few years, one can divide it into $N$ slots with a typical width of about one week, 
\begin{equation}
    d_{i}(t)=g_{i}(t)+n(t)+s(t)\;,
\end{equation}
where $g,n,s$ denote for the galactic foreground, gaussian stationary noise and isotropic SGWB signal, respectively. The sub-index labels the slot sort. 
Now, we weight the original data stream to reduce the foreground
\begin{equation}
    \tilde{d}(t)=\sum_{i}\omega_{i}d_{i}(t)\;, \quad\sum_{i}\omega_{i}=N\;,
\end{equation}
where the second equation ensures that the weighting will not damage SGWB strength.  We ask the weight to give the minimum variance of the periodic diagram. 
To do so, we give the power spectrum density (PSD) of the data stream $\tilde{d}(t)$
\begin{align}
    \tilde{S}(f) & = \langle |\tilde{d}(f)| ^{2}\rangle\;\nonumber\\
    & = \langle \frac{|\sum_{i}\omega_{i}g_{i}(f)| ^{2}}{N}\rangle + S_{n}(f)+S_{s}(f),\;\label{eq:psd}
\end{align}
where $\tilde{d}(f)$ is the fourier transform of the $\tilde{d}(t)$. One can easily see that the weight will not influence the PSD of instrument noise and the SGWB signal, which are stationary and gaussian during the observing campaign.

Using the precondition $\sum_{i}\omega_{i}=N$ and Lagrangian multiplier, one can derive the optimal weight as 
\begin{equation}\label{eq:IVF weight}
    \omega_{i}=N\frac{\Big[\int g^2_{i}(f)df\Big]^{-1}}{\sum_{j}\Big[\int g^2_{j}(f)df\Big]^{-1}}\;.
\end{equation}
One can see that this is simply the inverse variance filter. 
The behind logic is the following. On the one hand, the galactic foreground is anisotropic and follows the spatial distribution of the galactic plane. With the motion of the detector constellation, the galactic foreground is loud when the constellation points to the galactic plane; when the constellation points to the high galactic latitude, the galactic foreground is relatively low. On the other hand, the SGWB is isotropic. Hence, one can give a low weight to the former and a high weight to the latter.  

%%%%%%%%%%%%%%%%%%%%%
\begin{figure}[hb!]
\includegraphics[width = 8.6cm]{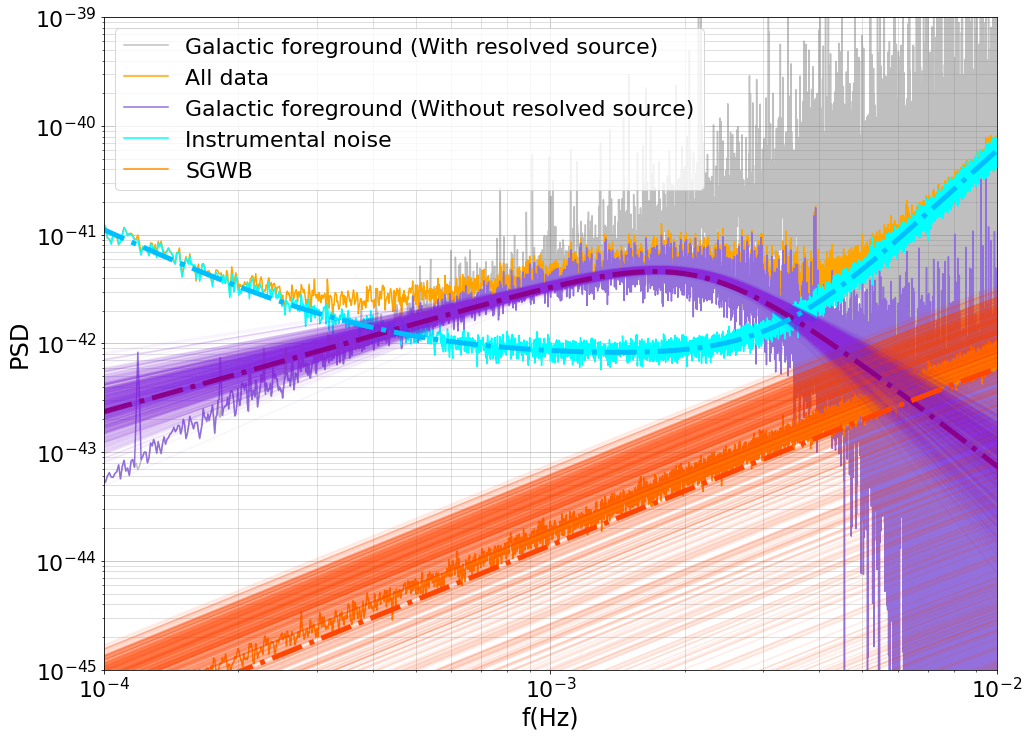}% Here is how to import EPS art
\caption{\label{fig:P3_PSD} PSD of the simulated components and the fitting results. The injected SGWB energy density reads $\Omega_{\rm astro}=1\times10^{-12}$.}
\end{figure}

%%%%%%%%%%%%%%%%%%%%%%%%%
\emph{Results.--} 
In the rest, we will use the LISA Data Challenge simulation to demonstrate the robustness of this method. 
LISA constellation consists of three spacecraft with a triangular configuration,  separated from one another at a distance of $2.5\times 10^{6}\; km$. 
To overcome the laser frequency noise, the LISA experiment adopts the time-delay interferometry (TDI) technology \cite{2002PhRvD..65h2003T}. 
In this Letter, we use the Michelson TDI configuration (X,Y,Z) to compose the optimal TDI configuration (A,E,T) \cite{2008CQGra..25f5005V}. 
For simplicity, we assume the equal arm configuration. Under these circumstances, the T-channel is a nearly null channel for GW signal search, but it can be used to calibrate the instrumental noise \cite{2010PhRvD..82b2002A}.
Following the same methodology presented in \cite{2019PhRvD.100j4055S}, we use the analytic expressions for the response functions and instrumental noise PSD. 
We consider two major components of the instrumental noise, namely the acceleration noise ($N_{\rm acc}$) and optical path-length fluctuation ($N_{\rm op}$). 

Furthermore, we adopt the galactic foreground model given by LDC1-4 data \cite{2010PhRvD..81f3008B}, which consists of about 30 million WDBs. 
For each binary, LDC1-4 contains  the amplitude, the frequency, the frequency derivative, the ecliptic latitude and longitude, the inclination, the initial phase and the polarization angle \cite{2021PhRvD.104b4023Z}. Among the 30 million WDBs, about ten thousand systems can be resolved individually. These strong point sources must be identified and removed accurately. 
To do so, we use the method from galactic binary separation by iterative extraction and validation using extended range (GBSIEVER) \cite{2021PhRvD.104b4023Z}. The principal novel features of GBSIEVER are using particle swarm optimization to maximize the F-statistic, fast template generation using under-sampling, and mitigation of spurious sources using a cross-validation scheme. By this means, we remove about ten thousand strong point sources. The initial PSD and the residual of the galactic foreground are shown as light-grey and light-purple curves in Fig.\ref{fig:P3_PSD}, respectively. One can clearly see that our foreground cleaning method significantly improves the results in the frequency range of ($10^{-3}-10^{-2}$) Hz. 

Finally, we consider the SGWB from the astrophysical origin(we also consider the cosmological model, details in appendix A), sourced by many compact binaries, mostly stellar-origin black holes and neutron star binaries (BBH+BNS). The gravitational wave emission of them is incoherently superposed. According to previous work \cite{2019RPPh...82a6903C}, this component can be well approximated by a power law function with a slope $\alpha_{\rm astro}=2/3$. In this Letter, we inject two amplitude values, namely $\Omega_{\rm astro}=1\times10^{-12}$ and $3\times10^{-12}$, respectively. 

Armed with these preparations, we simulate the time domain stream, $d(t)$, for two years length and with 15 seconds sampling rate. Then, we use the following three steps to process the IVF operation.
\begin{itemize}
    \item First, we divide the time domain data into 100 slots, $\{d_{i}(t)\}$, in which each slot has about 1 week time length. We also test if the selection of the slots will influence the result. Details are in appendix B.
    \item Second, we calculate the IVF for each slot with Eq.\eqref{eq:IVF weight}. The frequency integration range covers $[1\times10^{-4}-7 \times 10^{-3}]$Hz, where the galactic foregrounds are prominent.  
    \item At last, we obtain the filtered data, $\tilde{d}(t)$, by multiplying $\omega_{i}$ to the original data. Then, we calculate the PSD of the filtered data and estimate different components in the frequency domain. 
\end{itemize}

%%%%%%%%%%%%%%%%%%%%%
\begin{figure}[hb!]
\centering
\includegraphics[width = 8.6cm]{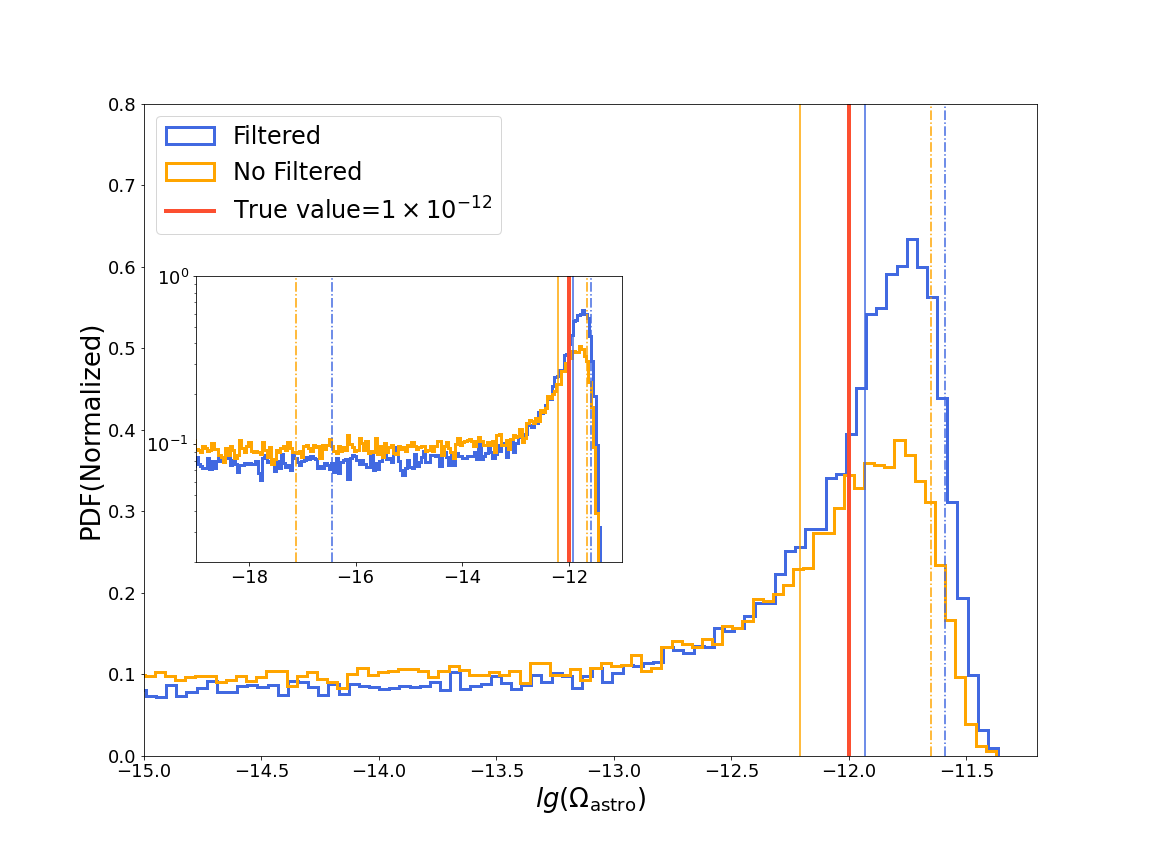}% Here is how to import EPS art
\caption{\label{fig:P1_1e-12} Posterior distribution function of SGWB energy density, $\Omega_{\rm astro}$, with (blue) and without (orange) the IVF. The injected value is $\Omega_{\rm astro}=1\times10^{-12}$. The solid and dashed vertical lines with the corresponding colors denote for the 1$\sigma$ and 2$\sigma$ ranges, respectively.}
\end{figure}

%%%%%%%%%%%%%%%%%%%%%
\begin{figure}[hb!]
\includegraphics[width = 8.6cm]{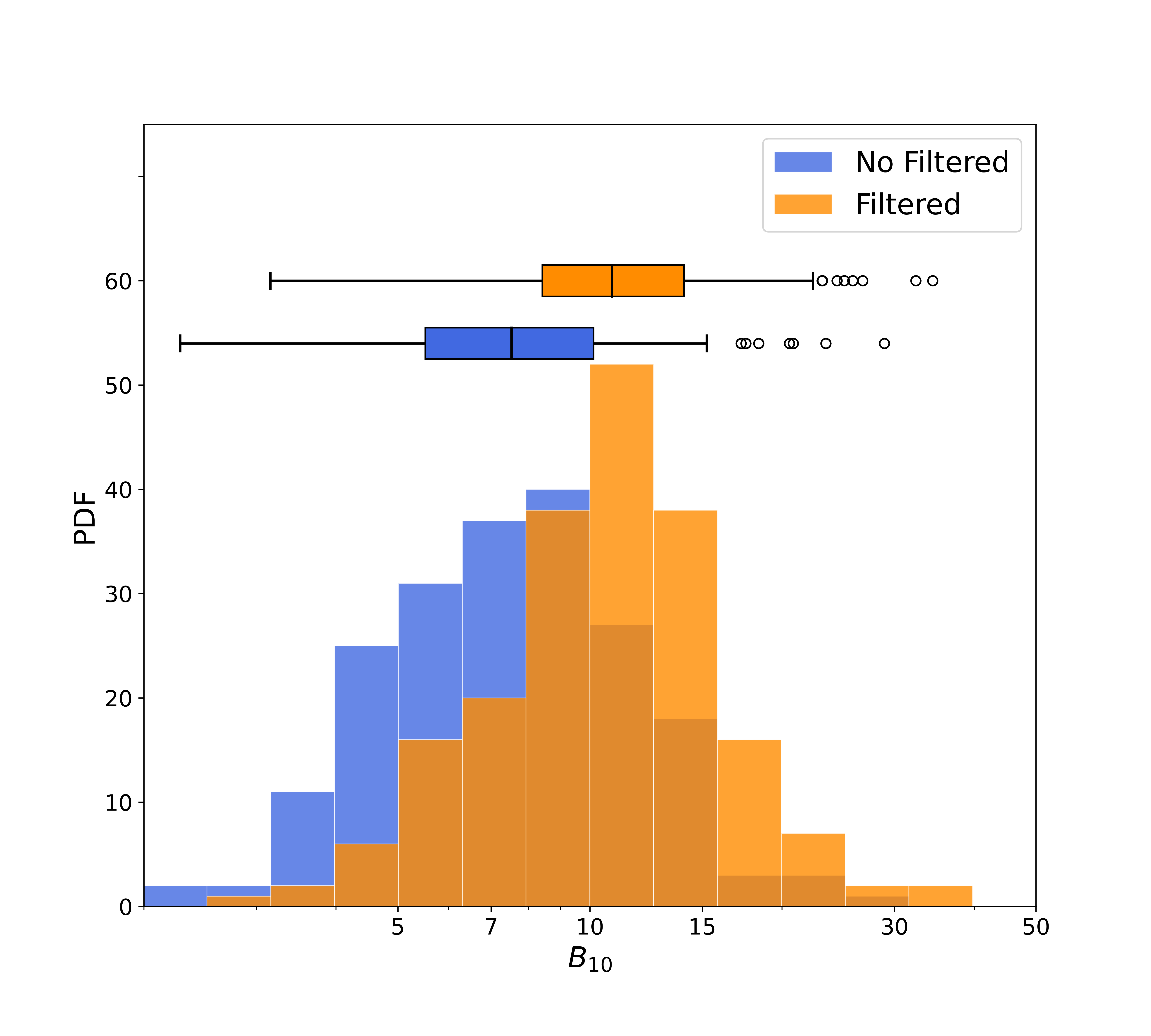}% Here is how to import EPS art
\caption{\label{fig:P2_3e-12} The probability distribution function of the Bayes factor with and without the IVF. The injected SGWB energy density value is $\Omega_{\rm astro}=3\times10^{-12}$. The number of the simulation is 200.}
\end{figure}

Next, we estimate the SGWB energy density, foregrounds as well as instrumental noise parameters via the Monte Carlo method. 
We write the PSD model as, $S_{n,I}+S_{{\rm gw},I}$, where $S_{n,I}$ is the instrumental noise PSD; and $S_{{\rm gw},I}$ denotes for the total GW energy density $\Omega_{\rm gw} = \Omega_{\rm fg}+\Omega_{\rm astro}$. 
To fit the galactic foreground, we use a broken power-law model \cite{2021MNRAS.508..803B}.
The number of the parameters in our analysis is seven ($N_{\rm pos},N_{\rm acc},A_{1},\alpha_{1},A_{2},\alpha_{2},\Omega_{\rm astro}$), two for instrument noise, four for galactic foregrounds and one for the SGWB amplitude.

 We use both the Markov chain Monte Carlo (MCMC) method to estimate the model parameters and dynamic nested sampling (DNS) to estimate Bayesian posteriors and evidences. The packages for MCMC and DNS are emcee \cite{2013PASP..125..306F} and dynesty \cite{2020MNRAS.493.3132S}, respectively. We adopt the Whittle likelihood \cite{2017LRR....20....2R}
\begin{equation} 
\begin{aligned}
L(\mathbf{d}|\theta)
=&-\dfrac{1}{2}\sum_{k=0}^{N}\Bigg[\dfrac{PSD_{A}}{S_{{\rm gw},A}+S_{n,A}}+\dfrac{PSD_{E}}{S_{{\rm gw},E}+S_{n,E}}+\dfrac{PSD_{T}}{S_{n,T}}\\
&+\log\Big\{8\pi^{3}(S_{{\rm gw},A}+S_{n,A})(S_{{\rm gw},E}+S_{n,E})S_{n,T}\Big\}\Bigg].
\end{aligned}
\end{equation}
Since the GW signal in the T-channel is much lower than the instrumental noise, we do not give the GW model in the T-channel but add it in our analysis to estimate the instrumental noise parameters. 

In Fig. \ref{fig:P3_PSD}, we show the fitting results from sampling points as the thin smoothing lines. Dark purple and orange transparent lines are for the foreground and SGWBs, respectively. The bold dashed curves denote for the best fit. In Fig. \ref{fig:P1_1e-12}, we plot the posterior of $\Omega_{\rm astro}$. The blue and orange curves denote for the results with and without the filtered method. The injected SGWB energy density is $\Omega_{\rm astro}=1\times10^{-12}$, marked as the red vertical line. Obviously, the filtered method makes the posterior peak more prominent. As demonstrated in the sub-panel, the posterior probability of $\Omega_{\rm astro}$ has a long tail extending to the lower values. After applying the IVF method, the left-hand tail gets depressed, and the corresponding masses will move to the right and contribute to the main peak.  
This benefit will also appear when we increase the amplitude of the SGWB. 

For a higher amplitude ($\Omega_{\rm astro}=3\times10^{-12}$), instead of the parameter estimation, we turn to the Bayesian model selection. 
We consider two models: Model $M_{0}$ only contains the instrument noise and galactic foregrounds; model $M_{1}$ has the instrumental noise, galactic foregrounds and SGWB. 
We simulated 200 realizations for each case with/without IVF to see the distribution of the Bayesian factor as shown in Fig. \ref{fig:P2_3e-12}. 
The blue and orange horizontal hinges denote for the quartiles, namely the $25\%-75\%$ confidence level. The tiny vertical line in the middle of the hinges denotes for the median value. The horizontal error bar ranges lie 1.5 times the inter-fourth range from the median. The empty circles denote for the outliers.
Without applying IVF, we obtain a Bayesian factor $B_{10,{\rm nof}}=7.5^{+2.6}_{-2.0}$; after applying IVF, we obtain a Bayesian factor $B_{10,{\rm f}}=10.8^{+3.2}_{-2.4}$. 
The corresponding difference statistics, $\Delta B_{10}= B_{10,{\rm f}}-B_{10,{\rm nof}}$, reads $\Delta B_{10}= 2.9_{-2.6}^{+2.6}$. 
One can see that the IVF operation can improve the Bayesian factor from `substantial', $B_{10}\in(3.2,10)$, to `strong', $B_{10}\in(10,100)$, against null hypotheses \cite{Kass1995BayesFA}.  
These results show that WDB modulation will significantly improve confidence in model selection. 

%%%%%%%%%%%%%%%%%%%%%%%%%
\emph{Conclusion.--} 
The search for the SGWB signal from astrophysical and cosmological origins is one of the major science cases for the spaceborne gravitational wave observatory. 
However, the presence of the galactic foreground will obstruct our road. 
Few works have discussed selecting tens of thousands of the resolved WDBs within the galactic foreground \cite{2007PhRvD..75d3008C,2010PhRvD..81f3008B,2018PhRvD..98d4029N,2020PhRvD.101l3021L,2021PhRvD.104d3019K,2021PhRvD.104b4023Z,2022PhRvD.106j2004Z,2022arXiv220502384L}. 
Some works have suggested that there would be confusion noise after removing each of the resolved WDBs and developed models of the ideal confusion noise \cite{2017JPhCS.840a2024C,2021PhRvD.104d3019K,2022ApJ...940...10D}. 
Some algorithms to separate different components have been proposed \cite{2014PhRvD..89b2001A,2020JCAP...07..021P,2021MNRAS.508..803B,2021JCAP...01..059F,2022PhRvD.106d4054W}. 
The annual modulation produced by the galactic WDBs can provide useful information to distinguish the galactic foreground from SGWBs. 

In this Letter, we propose a filtering method based on the annual modulation information. 
The advantage of this method is its model independency. 
We do not need to know how the WDBs are distributed in the galaxy, and we do not need to know the accurate position of the spaceborne GW interferometer. 
Every process is made on the observational gravitational wave data alone (simulated in this Letter) and no external data is needed. 
Besides, this method is independent of how well we select the resolved source. 
In this Letter, we only discuss the astrophysical originated SGWBs, described by a power-law function with the slope $\alpha=2/3$. 
This method will also be suitable for other forms of the SGWBs, like the flat spectrum for inflation \cite{2019RPPh...82a6903C}, the gaussian-bump spectrum for the first-order phase transition \cite{2022arXiv220913277B,2022arXiv220811615C} and the spectrum for the cosmic string \cite{2022PhRvD.105b3510B}. 
This idea does not depend on the amplitude of the SGWB signal. We show that it is well-behaved for both lower and higher amplitude. 
In conclusion, the annual modulation originated from WDBs can help stochastic gravitational wave background search.

%%%%%%%%%%%%%%%%%%%%%%%%%
\emph{Acknowledgements.--} 
This work is supported by the National Key R\&D Program of China No. 2021YFC2203001 and No. 2021YFC2203003, and the National Natural Science Foundation of China through Grant No. 12247101.

%%%%%%%%%%%%%%%%%%%%%%%%%
\appendix

\section{Bayesian model selection test for cosmological SGWB} 

\begin{figure}[hb!]
\includegraphics[width = 8.6cm]{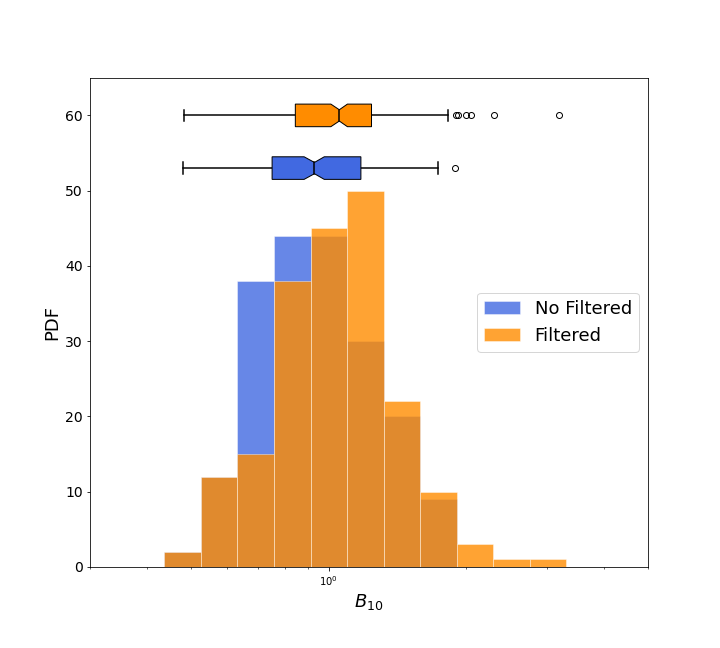}% Here is how to import EPS art
\caption{\label{fig:P2_7e-12} The probability distribution function of the Bayes factor with and without the IVF. The injected SGWB energy density value is $\Omega_{\rm astro}=7\times10^{-12}$ and the slope is $\alpha=0$. The number of the simulation is 200.}
\end{figure}
For the cosmological SGWB, we consider a flat spectrum that is predicted by inflation theory, and inject an amplitude value of $\Omega_{\rm cos}=7\times10^{-12}$.  The result is shown in Fig.\ref{fig:P2_7e-12}. Without applying IVF, we obtain a Bayesian factor $B_{10,{\rm nof}}=0.93^{+0.18}_{-0.24}$; after applying IVF, we obtain a Bayesian factor $B_{10,{\rm f}}=1.06^{+0.22}_{-0.16}$. The IVF operation can improve the Bayesian factor from `no evidence', $B_{10}<1$, to `littel evidence', $B_{10}\in(1,3.2)$, against null hypotheses. The result is similar to that obtained for an astrophysical origin SGWB, but it is not significant. This is because the flat spectrum predicted by the inflation theory is much lower than the astrophysical SGWB in the high-frequency range ($10^{-3}-10^{-2}, Hz$), which has a higher SNR (signal-to-noise ratio).

\section{Dependency of the number of slots}
\begin{figure}[hb!]
\includegraphics[width = 8.6cm]{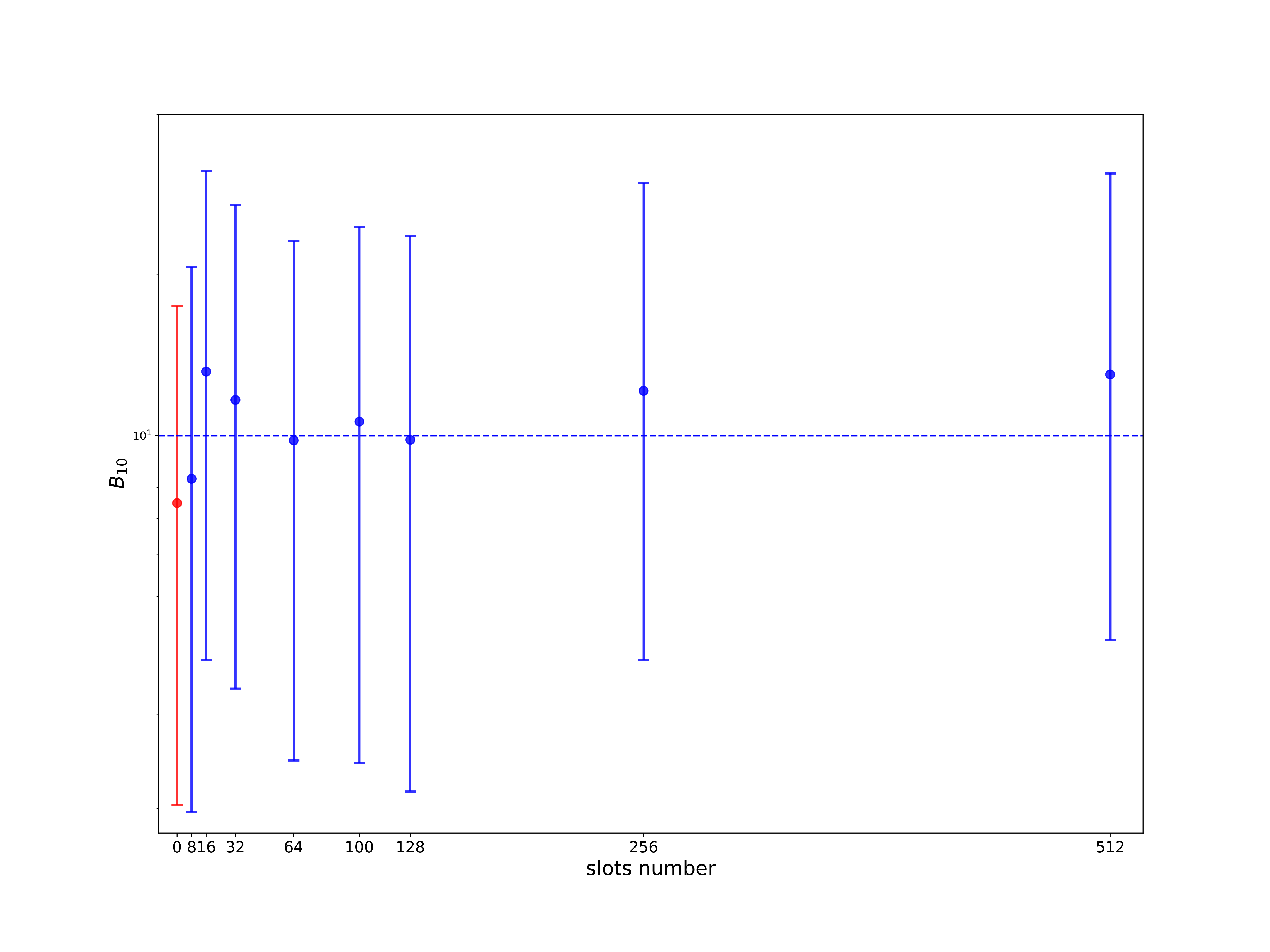}% Here is how to import EPS art
\caption{\label{fig:P4_slots} The dependence of the Bayesian factor on the slots number. The injected SGWB energy density value is $\Omega_{\rm astro}=3\times10^{-12}$ and the slope is $\alpha=2/3$. The first red point represent the without filtering case. The number of the simulation is 200 in each slots.}
\end{figure}
To investigate the dependence of our results on the choice of the number of slots, we varied the slot number from 8 (where each slot has a duration of approximately one season) to 512 (where each slot has a duration of approximately one day). We simulated 200 realizations for each slot number. The results in Fig. \ref{fig:P4_slots} indicate that for a large number of slots, the IVF operation can still improve the Bayesian factor from indicating "substantial" evidence ($B_{10}\in(3.2,10)$) to indicating "strong" evidence ($B_{10}\in(10,100)$) against null hypotheses.

\bibliography{apssamp}

\begin{thebibliography}{52}%
\makeatletter
\providecommand \@ifxundefined [1]{%
 \@ifx{#1\undefined}
}%
\providecommand \@ifnum [1]{%
 \ifnum #1\expandafter \@firstoftwo
 \else \expandafter \@secondoftwo
 \fi
}%
\providecommand \@ifx [1]{%
 \ifx #1\expandafter \@firstoftwo
 \else \expandafter \@secondoftwo
 \fi
}%
\providecommand \natexlab [1]{#1}%
\providecommand \enquote  [1]{``#1''}%
\providecommand \bibnamefont  [1]{#1}%
\providecommand \bibfnamefont [1]{#1}%
\providecommand \citenamefont [1]{#1}%
\providecommand \href@noop [0]{\@secondoftwo}%
\providecommand \href [0]{\begingroup \@sanitize@url \@href}%
\providecommand \@href[1]{\@@startlink{#1}\@@href}%
\providecommand \@@href[1]{\endgroup#1\@@endlink}%
\providecommand \@sanitize@url [0]{\catcode `\\12\catcode `\$12\catcode
  `\&12\catcode `\#12\catcode `\^12\catcode `\_12\catcode `\%12\relax}%
\providecommand \@@startlink[1]{}%
\providecommand \@@endlink[0]{}%
\providecommand \url  [0]{\begingroup\@sanitize@url \@url }%
\providecommand \@url [1]{\endgroup\@href {#1}{\urlprefix }}%
\providecommand \urlprefix  [0]{URL }%
\providecommand \Eprint [0]{\href }%
\providecommand \doibase [0]{http://dx.doi.org/}%
\providecommand \selectlanguage [0]{\@gobble}%
\providecommand \bibinfo  [0]{\@secondoftwo}%
\providecommand \bibfield  [0]{\@secondoftwo}%
\providecommand \translation [1]{[#1]}%
\providecommand \BibitemOpen [0]{}%
\providecommand \bibitemStop [0]{}%
\providecommand \bibitemNoStop [0]{.\EOS\space}%
\providecommand \EOS [0]{\spacefactor3000\relax}%
\providecommand \BibitemShut  [1]{\csname bibitem#1\endcsname}%
\let\auto@bib@innerbib\@empty
%</preamble>
\bibitem [{\citenamefont {{Allen}}\ and\ \citenamefont
  {{Romano}}(1999)}]{1999PhRvD..59j2001A}%
  \BibitemOpen
  \bibfield  {author} {\bibinfo {author} {\bibfnamefont {B.}~\bibnamefont
  {{Allen}}}\ and\ \bibinfo {author} {\bibfnamefont {J.~D.}\ \bibnamefont
  {{Romano}}},\ }\href {\doibase 10.1103/PhysRevD.59.102001} {\bibfield
  {journal} {\bibinfo  {journal} {\prd}\ }\textbf {\bibinfo {volume} {59}},\
  \bibinfo {eid} {102001} (\bibinfo {year} {1999})},\ \Eprint
  {http://arxiv.org/abs/gr-qc/9710117} {arXiv:gr-qc/9710117 [gr-qc]}
  \BibitemShut {NoStop}%
\bibitem [{\citenamefont {{Romano}}\ and\ \citenamefont
  {{Cornish}}(2017)}]{2017LRR....20....2R}%
  \BibitemOpen
  \bibfield  {author} {\bibinfo {author} {\bibfnamefont {J.~D.}\ \bibnamefont
  {{Romano}}}\ and\ \bibinfo {author} {\bibfnamefont {N.~J.}\ \bibnamefont
  {{Cornish}}},\ }\href {\doibase 10.1007/s41114-017-0004-1} {\bibfield
  {journal} {\bibinfo  {journal} {Living Reviews in Relativity}\ }\textbf
  {\bibinfo {volume} {20}},\ \bibinfo {eid} {2} (\bibinfo {year} {2017})},\
  \Eprint {http://arxiv.org/abs/1608.06889} {arXiv:1608.06889 [gr-qc]}
  \BibitemShut {NoStop}%
\bibitem [{\citenamefont {{Christensen}}(2019)}]{2019RPPh...82a6903C}%
  \BibitemOpen
  \bibfield  {author} {\bibinfo {author} {\bibfnamefont {N.}~\bibnamefont
  {{Christensen}}},\ }\href {\doibase 10.1088/1361-6633/aae6b5} {\bibfield
  {journal} {\bibinfo  {journal} {Reports on Progress in Physics}\ }\textbf
  {\bibinfo {volume} {82}},\ \bibinfo {eid} {016903} (\bibinfo {year}
  {2019})},\ \Eprint {http://arxiv.org/abs/1811.08797} {arXiv:1811.08797
  [gr-qc]} \BibitemShut {NoStop}%
\bibitem [{\citenamefont {{Abbott}}\ \emph {et~al.}(2016)\citenamefont
  {{Abbott}} \emph {et~al.}}]{2016PhRvL.116m1103A}%
  \BibitemOpen
  \bibfield  {author} {\bibinfo {author} {\bibfnamefont {B.~P.}\ \bibnamefont
  {{Abbott}}} \emph {et~al.} (\bibinfo {collaboration} {LIGO Scientific
  Collaboration and Virgo Collaboration}),\ }\href {\doibase
  10.1103/PhysRevLett.116.131103} {\bibfield  {journal} {\bibinfo  {journal}
  {\prl}\ }\textbf {\bibinfo {volume} {116}},\ \bibinfo {eid} {131103}
  (\bibinfo {year} {2016})},\ \Eprint {http://arxiv.org/abs/1602.03838}
  {arXiv:1602.03838 [gr-qc]} \BibitemShut {NoStop}%
\bibitem [{\citenamefont {{Abbott}}\ \emph {et~al.}(2018)\citenamefont
  {{Abbott}} \emph {et~al.}}]{2018PhRvL.120i1101A}%
  \BibitemOpen
  \bibfield  {author} {\bibinfo {author} {\bibfnamefont {B.~P.}\ \bibnamefont
  {{Abbott}}} \emph {et~al.} (\bibinfo {collaboration} {LIGO Scientific
  Collaboration and Virgo Collaboration}),\ }\href {\doibase
  10.1103/PhysRevLett.120.091101} {\bibfield  {journal} {\bibinfo  {journal}
  {\prl}\ }\textbf {\bibinfo {volume} {120}},\ \bibinfo {eid} {091101}
  (\bibinfo {year} {2018})},\ \Eprint {http://arxiv.org/abs/1710.05837}
  {arXiv:1710.05837 [gr-qc]} \BibitemShut {NoStop}%
\bibitem [{\citenamefont {{Regimbau}}(2011)}]{2011RAA....11..369R}%
  \BibitemOpen
  \bibfield  {author} {\bibinfo {author} {\bibfnamefont {T.}~\bibnamefont
  {{Regimbau}}},\ }\href {\doibase 10.1088/1674-4527/11/4/001} {\bibfield
  {journal} {\bibinfo  {journal} {Research in Astronomy and Astrophysics}\
  }\textbf {\bibinfo {volume} {11}},\ \bibinfo {pages} {369} (\bibinfo {year}
  {2011})},\ \Eprint {http://arxiv.org/abs/1101.2762} {arXiv:1101.2762
  [astro-ph.CO]} \BibitemShut {NoStop}%
\bibitem [{\citenamefont {{Chen}}\ \emph {et~al.}(2019)\citenamefont {{Chen}},
  \citenamefont {{Huang}},\ and\ \citenamefont
  {{Huang}}}]{2019ApJ...871...97C}%
  \BibitemOpen
  \bibfield  {author} {\bibinfo {author} {\bibfnamefont {Z.-C.}\ \bibnamefont
  {{Chen}}}, \bibinfo {author} {\bibfnamefont {F.}~\bibnamefont {{Huang}}}, \
  and\ \bibinfo {author} {\bibfnamefont {Q.-G.}\ \bibnamefont {{Huang}}},\
  }\href {\doibase 10.3847/1538-4357/aaf581} {\bibfield  {journal} {\bibinfo
  {journal} {\apj}\ }\textbf {\bibinfo {volume} {871}},\ \bibinfo {eid} {97}
  (\bibinfo {year} {2019})},\ \Eprint {http://arxiv.org/abs/1809.10360}
  {arXiv:1809.10360 [gr-qc]} \BibitemShut {NoStop}%
\bibitem [{\citenamefont {{Bin{\'e}truy}}\ \emph {et~al.}(2012)\citenamefont
  {{Bin{\'e}truy}}, \citenamefont {{Boh{\'e}}}, \citenamefont {{Caprini}},\
  and\ \citenamefont {{Dufaux}}}]{2012JCAP...06..027B}%
  \BibitemOpen
  \bibfield  {author} {\bibinfo {author} {\bibfnamefont {P.}~\bibnamefont
  {{Bin{\'e}truy}}}, \bibinfo {author} {\bibfnamefont {A.}~\bibnamefont
  {{Boh{\'e}}}}, \bibinfo {author} {\bibfnamefont {C.}~\bibnamefont
  {{Caprini}}}, \ and\ \bibinfo {author} {\bibfnamefont {J.-F.}\ \bibnamefont
  {{Dufaux}}},\ }\href {\doibase 10.1088/1475-7516/2012/06/027} {\bibfield
  {journal} {\bibinfo  {journal} {\jcap}\ }\textbf {\bibinfo {volume} {2012}},\
  \bibinfo {eid} {027} (\bibinfo {year} {2012})},\ \Eprint
  {http://arxiv.org/abs/1201.0983} {arXiv:1201.0983 [gr-qc]} \BibitemShut
  {NoStop}%
\bibitem [{\citenamefont {{Caprini}}\ and\ \citenamefont
  {{Figueroa}}(2018)}]{2018CQGra..35p3001C}%
  \BibitemOpen
  \bibfield  {author} {\bibinfo {author} {\bibfnamefont {C.}~\bibnamefont
  {{Caprini}}}\ and\ \bibinfo {author} {\bibfnamefont {D.~G.}\ \bibnamefont
  {{Figueroa}}},\ }\href {\doibase 10.1088/1361-6382/aac608} {\bibfield
  {journal} {\bibinfo  {journal} {Classical and Quantum Gravity}\ }\textbf
  {\bibinfo {volume} {35}},\ \bibinfo {eid} {163001} (\bibinfo {year}
  {2018})},\ \Eprint {http://arxiv.org/abs/1801.04268} {arXiv:1801.04268
  [astro-ph.CO]} \BibitemShut {NoStop}%
\bibitem [{\citenamefont {{Abbott}}\ \emph {et~al.}(2017)\citenamefont
  {{Abbott}} \emph {et~al.}}]{2017PhRvL.118l1101A}%
  \BibitemOpen
  \bibfield  {author} {\bibinfo {author} {\bibfnamefont {B.~P.}\ \bibnamefont
  {{Abbott}}} \emph {et~al.} (\bibinfo {collaboration} {LIGO Scientific
  Collaboration and Virgo Collaboration}),\ }\href {\doibase
  10.1103/PhysRevLett.118.121101} {\bibfield  {journal} {\bibinfo  {journal}
  {\prl}\ }\textbf {\bibinfo {volume} {118}},\ \bibinfo {eid} {121101}
  (\bibinfo {year} {2017})},\ \Eprint {http://arxiv.org/abs/1612.02029}
  {arXiv:1612.02029 [gr-qc]} \BibitemShut {NoStop}%
\bibitem [{\citenamefont {{Abbott}}\ \emph {et~al.}(2019)\citenamefont
  {{Abbott}} \emph {et~al.}}]{2019PhRvD.100f1101A}%
  \BibitemOpen
  \bibfield  {author} {\bibinfo {author} {\bibfnamefont {B.~P.}\ \bibnamefont
  {{Abbott}}} \emph {et~al.} (\bibinfo {collaboration} {LIGO Scientific
  Collaboration and Virgo Collaboration}),\ }\href {\doibase
  10.1103/PhysRevD.100.061101} {\bibfield  {journal} {\bibinfo  {journal}
  {\prd}\ }\textbf {\bibinfo {volume} {100}},\ \bibinfo {eid} {061101}
  (\bibinfo {year} {2019})},\ \Eprint {http://arxiv.org/abs/1903.02886}
  {arXiv:1903.02886 [gr-qc]} \BibitemShut {NoStop}%
\bibitem [{\citenamefont {{Abbott}}\ \emph {et~al.}(2021)\citenamefont
  {{Abbott}} \emph {et~al.}}]{2021PhRvD.104b2004A}%
  \BibitemOpen
  \bibfield  {author} {\bibinfo {author} {\bibfnamefont {R.}~\bibnamefont
  {{Abbott}}} \emph {et~al.} (\bibinfo {collaboration} {LIGO Scientific
  Collaboration, Virgo Collaboration and KAGRA Collaboration}),\ }\href
  {\doibase 10.1103/PhysRevD.104.022004} {\bibfield  {journal} {\bibinfo
  {journal} {\prd}\ }\textbf {\bibinfo {volume} {104}},\ \bibinfo {eid}
  {022004} (\bibinfo {year} {2021})},\ \Eprint
  {http://arxiv.org/abs/2101.12130} {arXiv:2101.12130 [gr-qc]} \BibitemShut
  {NoStop}%
\bibitem [{\citenamefont {{Arzoumanian}}\ \emph {et~al.}(2020)\citenamefont
  {{Arzoumanian}} \emph {et~al.}}]{2020ApJ...905L..34A}%
  \BibitemOpen
  \bibfield  {author} {\bibinfo {author} {\bibfnamefont {Z.}~\bibnamefont
  {{Arzoumanian}}} \emph {et~al.} (\bibinfo {collaboration} {NANOGrav
  Collabration}),\ }\href {\doibase 10.3847/2041-8213/abd401} {\bibfield
  {journal} {\bibinfo  {journal} {\apjl}\ }\textbf {\bibinfo {volume} {905}},\
  \bibinfo {eid} {L34} (\bibinfo {year} {2020})},\ \Eprint
  {http://arxiv.org/abs/2009.04496} {arXiv:2009.04496 [astro-ph.HE]}
  \BibitemShut {NoStop}%
\bibitem [{\citenamefont {{Amaro-Seoane}}\ \emph {et~al.}()\citenamefont
  {{Amaro-Seoane}} \emph {et~al.}}]{2017arXiv170200786A}%
  \BibitemOpen
  \bibfield  {author} {\bibinfo {author} {\bibfnamefont {P.}~\bibnamefont
  {{Amaro-Seoane}}} \emph {et~al.} (\bibinfo {collaboration} {LISA
  Collaboration}),\ }\href@noop {} {\ }\Eprint {http://arxiv.org/abs/1702.00786
  (2017)} {arXiv:1702.00786 (2017) [astro-ph.IM]} \BibitemShut {NoStop}%
\bibitem [{\citenamefont {{Ruan}}\ \emph {et~al.}(2020)\citenamefont {{Ruan}},
  \citenamefont {{Guo}}, \citenamefont {{Cai}},\ and\ \citenamefont
  {{Zhang}}}]{2020IJMPA..3550075R}%
  \BibitemOpen
  \bibfield  {author} {\bibinfo {author} {\bibfnamefont {W.-H.}\ \bibnamefont
  {{Ruan}}}, \bibinfo {author} {\bibfnamefont {Z.-K.}\ \bibnamefont {{Guo}}},
  \bibinfo {author} {\bibfnamefont {R.-G.}\ \bibnamefont {{Cai}}}, \ and\
  \bibinfo {author} {\bibfnamefont {Y.-Z.}\ \bibnamefont {{Zhang}}},\ }\href
  {\doibase 10.1142/S0217751X2050075X} {\bibfield  {journal} {\bibinfo
  {journal} {International Journal of Modern Physics A}\ }\textbf {\bibinfo
  {volume} {35}},\ \bibinfo {eid} {2050075} (\bibinfo {year}
  {2020})}\BibitemShut {NoStop}%
\bibitem [{\citenamefont {{Luo}}\ \emph {et~al.}(2016)\citenamefont {{Luo}}
  \emph {et~al.}}]{2016CQGra..33c5010L}%
  \BibitemOpen
  \bibfield  {author} {\bibinfo {author} {\bibfnamefont {J.}~\bibnamefont
  {{Luo}}} \emph {et~al.} (\bibinfo {collaboration} {TianQin Collaboration}),\
  }\href {\doibase 10.1088/0264-9381/33/3/035010} {\bibfield  {journal}
  {\bibinfo  {journal} {Classical and Quantum Gravity}\ }\textbf {\bibinfo
  {volume} {33}},\ \bibinfo {eid} {035010} (\bibinfo {year} {2016})},\ \Eprint
  {http://arxiv.org/abs/1512.02076} {arXiv:1512.02076 [astro-ph.IM]}
  \BibitemShut {NoStop}%
\bibitem [{\citenamefont {{Zhao}}\ and\ \citenamefont
  {{Lu}}(2021)}]{2021MNRAS.500.1421Z}%
  \BibitemOpen
  \bibfield  {author} {\bibinfo {author} {\bibfnamefont {Y.}~\bibnamefont
  {{Zhao}}}\ and\ \bibinfo {author} {\bibfnamefont {Y.}~\bibnamefont {{Lu}}},\
  }\href {\doibase 10.1093/mnras/staa2707} {\bibfield  {journal} {\bibinfo
  {journal} {\mnras}\ }\textbf {\bibinfo {volume} {500}},\ \bibinfo {pages}
  {1421} (\bibinfo {year} {2021})},\ \Eprint {http://arxiv.org/abs/2009.01436}
  {arXiv:2009.01436 [astro-ph.HE]} \BibitemShut {NoStop}%
\bibitem [{\citenamefont {{Evans}}\ \emph {et~al.}(1987)\citenamefont
  {{Evans}}, \citenamefont {{Iben}},\ and\ \citenamefont
  {{Smarr}}}]{1987ApJ...323..129E}%
  \BibitemOpen
  \bibfield  {author} {\bibinfo {author} {\bibfnamefont {C.~R.}\ \bibnamefont
  {{Evans}}}, \bibinfo {author} {\bibfnamefont {I.}~\bibnamefont {{Iben}}}, \
  and\ \bibinfo {author} {\bibfnamefont {L.}~\bibnamefont {{Smarr}}},\ }\href
  {\doibase 10.1086/165812} {\bibfield  {journal} {\bibinfo  {journal} {\apj}\
  }\textbf {\bibinfo {volume} {323}},\ \bibinfo {pages} {129} (\bibinfo {year}
  {1987})}\BibitemShut {NoStop}%
\bibitem [{\citenamefont {{Bender}}\ and\ \citenamefont
  {{Hils}}(1997)}]{1997CQGra..14.1439B}%
  \BibitemOpen
  \bibfield  {author} {\bibinfo {author} {\bibfnamefont {P.~L.}\ \bibnamefont
  {{Bender}}}\ and\ \bibinfo {author} {\bibfnamefont {D.}~\bibnamefont
  {{Hils}}},\ }\href {\doibase 10.1088/0264-9381/14/6/008} {\bibfield
  {journal} {\bibinfo  {journal} {Classical and Quantum Gravity}\ }\textbf
  {\bibinfo {volume} {14}},\ \bibinfo {pages} {1439} (\bibinfo {year}
  {1997})}\BibitemShut {NoStop}%
\bibitem [{\citenamefont {{Nelemans}}\ \emph {et~al.}(2001)\citenamefont
  {{Nelemans}}, \citenamefont {{Yungelson}},\ and\ \citenamefont {{Portegies
  Zwart}}}]{2001A&A...375..890N}%
  \BibitemOpen
  \bibfield  {author} {\bibinfo {author} {\bibfnamefont {G.}~\bibnamefont
  {{Nelemans}}}, \bibinfo {author} {\bibfnamefont {L.~R.}\ \bibnamefont
  {{Yungelson}}}, \ and\ \bibinfo {author} {\bibfnamefont {S.~F.}\ \bibnamefont
  {{Portegies Zwart}}},\ }\href {\doibase 10.1051/0004-6361:20010683}
  {\bibfield  {journal} {\bibinfo  {journal} {\aap}\ }\textbf {\bibinfo
  {volume} {375}},\ \bibinfo {pages} {890} (\bibinfo {year} {2001})},\ \Eprint
  {http://arxiv.org/abs/astro-ph/0105221} {arXiv:astro-ph/0105221 [astro-ph]}
  \BibitemShut {NoStop}%
\bibitem [{\citenamefont {{Timpano}}\ \emph {et~al.}(2006)\citenamefont
  {{Timpano}}, \citenamefont {{Rubbo}},\ and\ \citenamefont
  {{Cornish}}}]{2006PhRvD..73l2001T}%
  \BibitemOpen
  \bibfield  {author} {\bibinfo {author} {\bibfnamefont {S.~E.}\ \bibnamefont
  {{Timpano}}}, \bibinfo {author} {\bibfnamefont {L.~J.}\ \bibnamefont
  {{Rubbo}}}, \ and\ \bibinfo {author} {\bibfnamefont {N.~J.}\ \bibnamefont
  {{Cornish}}},\ }\href {\doibase 10.1103/PhysRevD.73.122001} {\bibfield
  {journal} {\bibinfo  {journal} {\prd}\ }\textbf {\bibinfo {volume} {73}},\
  \bibinfo {eid} {122001} (\bibinfo {year} {2006})},\ \Eprint
  {http://arxiv.org/abs/gr-qc/0504071} {arXiv:gr-qc/0504071 [gr-qc]}
  \BibitemShut {NoStop}%
\bibitem [{\citenamefont {{Babak}}\ \emph {et~al.}(2008)\citenamefont {{Babak}}
  \emph {et~al.}}]{2008CQGra..25r4026B}%
  \BibitemOpen
  \bibfield  {author} {\bibinfo {author} {\bibfnamefont {S.}~\bibnamefont
  {{Babak}}} \emph {et~al.},\ }\href {\doibase 10.1088/0264-9381/25/18/184026}
  {\bibfield  {journal} {\bibinfo  {journal} {Classical and Quantum Gravity}\
  }\textbf {\bibinfo {volume} {25}},\ \bibinfo {eid} {184026} (\bibinfo {year}
  {2008})},\ \Eprint {http://arxiv.org/abs/0806.2110} {arXiv:0806.2110 [gr-qc]}
  \BibitemShut {NoStop}%
\bibitem [{\citenamefont {{Adams}}\ and\ \citenamefont
  {{Cornish}}(2014)}]{2014PhRvD..89b2001A}%
  \BibitemOpen
  \bibfield  {author} {\bibinfo {author} {\bibfnamefont {M.~R.}\ \bibnamefont
  {{Adams}}}\ and\ \bibinfo {author} {\bibfnamefont {N.~J.}\ \bibnamefont
  {{Cornish}}},\ }\href {\doibase 10.1103/PhysRevD.89.022001} {\bibfield
  {journal} {\bibinfo  {journal} {\prd}\ }\textbf {\bibinfo {volume} {89}},\
  \bibinfo {eid} {022001} (\bibinfo {year} {2014})},\ \Eprint
  {http://arxiv.org/abs/1307.4116} {arXiv:1307.4116 [gr-qc]} \BibitemShut
  {NoStop}%
\bibitem [{\citenamefont {{Flauger}}\ \emph {et~al.}(2021)\citenamefont
  {{Flauger}}, \citenamefont {{Karnesis}}, \citenamefont {{Nardini}},
  \citenamefont {{Pieroni}}, \citenamefont {{Ricciardone}},\ and\ \citenamefont
  {{Torrado}}}]{2021JCAP...01..059F}%
  \BibitemOpen
  \bibfield  {author} {\bibinfo {author} {\bibfnamefont {R.}~\bibnamefont
  {{Flauger}}}, \bibinfo {author} {\bibfnamefont {N.}~\bibnamefont
  {{Karnesis}}}, \bibinfo {author} {\bibfnamefont {G.}~\bibnamefont
  {{Nardini}}}, \bibinfo {author} {\bibfnamefont {M.}~\bibnamefont
  {{Pieroni}}}, \bibinfo {author} {\bibfnamefont {A.}~\bibnamefont
  {{Ricciardone}}}, \ and\ \bibinfo {author} {\bibfnamefont {J.}~\bibnamefont
  {{Torrado}}},\ }\href {\doibase 10.1088/1475-7516/2021/01/059} {\bibfield
  {journal} {\bibinfo  {journal} {\jcap}\ }\textbf {\bibinfo {volume} {2021}},\
  \bibinfo {eid} {059} (\bibinfo {year} {2021})},\ \Eprint
  {http://arxiv.org/abs/2009.11845} {arXiv:2009.11845 [astro-ph.CO]}
  \BibitemShut {NoStop}%
\bibitem [{\citenamefont {{Poletti}}(2021)}]{2021JCAP...05..052P}%
  \BibitemOpen
  \bibfield  {author} {\bibinfo {author} {\bibfnamefont {D.}~\bibnamefont
  {{Poletti}}},\ }\href {\doibase 10.1088/1475-7516/2021/05/052} {\bibfield
  {journal} {\bibinfo  {journal} {\jcap}\ }\textbf {\bibinfo {volume} {2021}},\
  \bibinfo {eid} {052} (\bibinfo {year} {2021})},\ \Eprint
  {http://arxiv.org/abs/2101.02713} {arXiv:2101.02713 [gr-qc]} \BibitemShut
  {NoStop}%
\bibitem [{\citenamefont {{Crowder}}\ and\ \citenamefont
  {{Cornish}}(2007)}]{2007PhRvD..75d3008C}%
  \BibitemOpen
  \bibfield  {author} {\bibinfo {author} {\bibfnamefont {J.}~\bibnamefont
  {{Crowder}}}\ and\ \bibinfo {author} {\bibfnamefont {N.~J.}\ \bibnamefont
  {{Cornish}}},\ }\href {\doibase 10.1103/PhysRevD.75.043008} {\bibfield
  {journal} {\bibinfo  {journal} {\prd}\ }\textbf {\bibinfo {volume} {75}},\
  \bibinfo {eid} {043008} (\bibinfo {year} {2007})},\ \Eprint
  {http://arxiv.org/abs/astro-ph/0611546} {arXiv:astro-ph/0611546 [astro-ph]}
  \BibitemShut {NoStop}%
\bibitem [{\citenamefont {{Zhang}}\ \emph {et~al.}(2021)\citenamefont
  {{Zhang}}, \citenamefont {{Mohanty}}, \citenamefont {{Zou}},\ and\
  \citenamefont {{Liu}}}]{2021PhRvD.104b4023Z}%
  \BibitemOpen
  \bibfield  {author} {\bibinfo {author} {\bibfnamefont {X.-H.}\ \bibnamefont
  {{Zhang}}}, \bibinfo {author} {\bibfnamefont {S.~D.}\ \bibnamefont
  {{Mohanty}}}, \bibinfo {author} {\bibfnamefont {X.-B.}\ \bibnamefont
  {{Zou}}}, \ and\ \bibinfo {author} {\bibfnamefont {Y.-X.}\ \bibnamefont
  {{Liu}}},\ }\href {\doibase 10.1103/PhysRevD.104.024023} {\bibfield
  {journal} {\bibinfo  {journal} {\prd}\ }\textbf {\bibinfo {volume} {104}},\
  \bibinfo {eid} {024023} (\bibinfo {year} {2021})},\ \Eprint
  {http://arxiv.org/abs/2103.09391} {arXiv:2103.09391 [gr-qc]} \BibitemShut
  {NoStop}%
\bibitem [{\citenamefont {{Lamberts}}\ \emph {et~al.}(2019)\citenamefont
  {{Lamberts}}, \citenamefont {{Blunt}}, \citenamefont {{Littenberg}},
  \citenamefont {{Garrison-Kimmel}}, \citenamefont {{Kupfer}},\ and\
  \citenamefont {{Sanderson}}}]{2019MNRAS.490.5888L}%
  \BibitemOpen
  \bibfield  {author} {\bibinfo {author} {\bibfnamefont {A.}~\bibnamefont
  {{Lamberts}}}, \bibinfo {author} {\bibfnamefont {S.}~\bibnamefont {{Blunt}}},
  \bibinfo {author} {\bibfnamefont {T.~B.}\ \bibnamefont {{Littenberg}}},
  \bibinfo {author} {\bibfnamefont {S.}~\bibnamefont {{Garrison-Kimmel}}},
  \bibinfo {author} {\bibfnamefont {T.}~\bibnamefont {{Kupfer}}}, \ and\
  \bibinfo {author} {\bibfnamefont {R.~E.}\ \bibnamefont {{Sanderson}}},\
  }\href {\doibase 10.1093/mnras/stz2834} {\bibfield  {journal} {\bibinfo
  {journal} {\mnras}\ }\textbf {\bibinfo {volume} {490}},\ \bibinfo {pages}
  {5888} (\bibinfo {year} {2019})},\ \Eprint {http://arxiv.org/abs/1907.00014}
  {arXiv:1907.00014 [astro-ph.HE]} \BibitemShut {NoStop}%
\bibitem [{\citenamefont {{Seto}}(2004)}]{2004PhRvD..69l3005S}%
  \BibitemOpen
  \bibfield  {author} {\bibinfo {author} {\bibfnamefont {N.}~\bibnamefont
  {{Seto}}},\ }\href {\doibase 10.1103/PhysRevD.69.123005} {\bibfield
  {journal} {\bibinfo  {journal} {\prd}\ }\textbf {\bibinfo {volume} {69}},\
  \bibinfo {eid} {123005} (\bibinfo {year} {2004})},\ \Eprint
  {http://arxiv.org/abs/gr-qc/0403014} {arXiv:gr-qc/0403014 [gr-qc]}
  \BibitemShut {NoStop}%
\bibitem [{\citenamefont {{Boileau}}\ \emph {et~al.}(2021)\citenamefont
  {{Boileau}}, \citenamefont {{Lamberts}}, \citenamefont {{Christensen}},
  \citenamefont {{Cornish}},\ and\ \citenamefont
  {{Meyer}}}]{2021MNRAS.508..803B}%
  \BibitemOpen
  \bibfield  {author} {\bibinfo {author} {\bibfnamefont {G.}~\bibnamefont
  {{Boileau}}}, \bibinfo {author} {\bibfnamefont {A.}~\bibnamefont
  {{Lamberts}}}, \bibinfo {author} {\bibfnamefont {N.}~\bibnamefont
  {{Christensen}}}, \bibinfo {author} {\bibfnamefont {N.~J.}\ \bibnamefont
  {{Cornish}}}, \ and\ \bibinfo {author} {\bibfnamefont {R.}~\bibnamefont
  {{Meyer}}},\ }\href {\doibase 10.1093/mnras/stab2575} {\bibfield  {journal}
  {\bibinfo  {journal} {\mnras}\ }\textbf {\bibinfo {volume} {508}},\ \bibinfo
  {pages} {803} (\bibinfo {year} {2021})},\ \Eprint
  {http://arxiv.org/abs/2105.04283} {arXiv:2105.04283 [gr-qc]} \BibitemShut
  {NoStop}%
\bibitem [{\citenamefont {{Hinshaw}}\ \emph {et~al.}(2007)\citenamefont
  {{Hinshaw}} \emph {et~al.}}]{2007ApJS..170..288H}%
  \BibitemOpen
  \bibfield  {author} {\bibinfo {author} {\bibfnamefont {G.}~\bibnamefont
  {{Hinshaw}}} \emph {et~al.} (\bibinfo {collaboration} {WMAP Collaboration}),\
  }\href {\doibase 10.1086/513698} {\bibfield  {journal} {\bibinfo  {journal}
  {\apjs}\ }\textbf {\bibinfo {volume} {170}},\ \bibinfo {pages} {288}
  (\bibinfo {year} {2007})},\ \Eprint {http://arxiv.org/abs/astro-ph/0603451}
  {arXiv:astro-ph/0603451 [astro-ph]} \BibitemShut {NoStop}%
\bibitem [{\citenamefont {{Carron}}\ and\ \citenamefont
  {{Lewis}}(2017)}]{2017PhRvD..96f3510C}%
  \BibitemOpen
  \bibfield  {author} {\bibinfo {author} {\bibfnamefont {J.}~\bibnamefont
  {{Carron}}}\ and\ \bibinfo {author} {\bibfnamefont {A.}~\bibnamefont
  {{Lewis}}},\ }\href {\doibase 10.1103/PhysRevD.96.063510} {\bibfield
  {journal} {\bibinfo  {journal} {\prd}\ }\textbf {\bibinfo {volume} {96}},\
  \bibinfo {eid} {063510} (\bibinfo {year} {2017})},\ \Eprint
  {http://arxiv.org/abs/1704.08230} {arXiv:1704.08230 [astro-ph.CO]}
  \BibitemShut {NoStop}%
\bibitem [{\citenamefont {{Tinto}}\ \emph {et~al.}(2002)\citenamefont
  {{Tinto}}, \citenamefont {{Estabrook}},\ and\ \citenamefont
  {{Armstrong}}}]{2002PhRvD..65h2003T}%
  \BibitemOpen
  \bibfield  {author} {\bibinfo {author} {\bibfnamefont {M.}~\bibnamefont
  {{Tinto}}}, \bibinfo {author} {\bibfnamefont {F.~B.}\ \bibnamefont
  {{Estabrook}}}, \ and\ \bibinfo {author} {\bibfnamefont {J.~W.}\ \bibnamefont
  {{Armstrong}}},\ }\href {\doibase 10.1103/PhysRevD.65.082003} {\bibfield
  {journal} {\bibinfo  {journal} {\prd}\ }\textbf {\bibinfo {volume} {65}},\
  \bibinfo {eid} {082003} (\bibinfo {year} {2002})}\BibitemShut {NoStop}%
\bibitem [{\citenamefont {{Vallisneri}}\ \emph {et~al.}(2008)\citenamefont
  {{Vallisneri}}, \citenamefont {{Crowder}},\ and\ \citenamefont
  {{Tinto}}}]{2008CQGra..25f5005V}%
  \BibitemOpen
  \bibfield  {author} {\bibinfo {author} {\bibfnamefont {M.}~\bibnamefont
  {{Vallisneri}}}, \bibinfo {author} {\bibfnamefont {J.}~\bibnamefont
  {{Crowder}}}, \ and\ \bibinfo {author} {\bibfnamefont {M.}~\bibnamefont
  {{Tinto}}},\ }\href {\doibase 10.1088/0264-9381/25/6/065005} {\bibfield
  {journal} {\bibinfo  {journal} {Classical and Quantum Gravity}\ }\textbf
  {\bibinfo {volume} {25}},\ \bibinfo {eid} {065005} (\bibinfo {year}
  {2008})},\ \Eprint {http://arxiv.org/abs/0710.4369} {arXiv:0710.4369 [gr-qc]}
  \BibitemShut {NoStop}%
\bibitem [{\citenamefont {{Adams}}\ and\ \citenamefont
  {{Cornish}}(2010)}]{2010PhRvD..82b2002A}%
  \BibitemOpen
  \bibfield  {author} {\bibinfo {author} {\bibfnamefont {M.~R.}\ \bibnamefont
  {{Adams}}}\ and\ \bibinfo {author} {\bibfnamefont {N.~J.}\ \bibnamefont
  {{Cornish}}},\ }\href {\doibase 10.1103/PhysRevD.82.022002} {\bibfield
  {journal} {\bibinfo  {journal} {\prd}\ }\textbf {\bibinfo {volume} {82}},\
  \bibinfo {eid} {022002} (\bibinfo {year} {2010})},\ \Eprint
  {http://arxiv.org/abs/1002.1291} {arXiv:1002.1291 [gr-qc]} \BibitemShut
  {NoStop}%
\bibitem [{\citenamefont {{Smith}}\ and\ \citenamefont
  {{Caldwell}}(2019)}]{2019PhRvD.100j4055S}%
  \BibitemOpen
  \bibfield  {author} {\bibinfo {author} {\bibfnamefont {T.~L.}\ \bibnamefont
  {{Smith}}}\ and\ \bibinfo {author} {\bibfnamefont {R.~R.}\ \bibnamefont
  {{Caldwell}}},\ }\href {\doibase 10.1103/PhysRevD.100.104055} {\bibfield
  {journal} {\bibinfo  {journal} {\prd}\ }\textbf {\bibinfo {volume} {100}},\
  \bibinfo {eid} {104055} (\bibinfo {year} {2019})},\ \Eprint
  {http://arxiv.org/abs/1908.00546} {arXiv:1908.00546 [astro-ph.CO]}
  \BibitemShut {NoStop}%
\bibitem [{\citenamefont {{B{\l}aut}}\ \emph {et~al.}(2010)\citenamefont
  {{B{\l}aut}}, \citenamefont {{Babak}},\ and\ \citenamefont
  {{Kr{\'o}lak}}}]{2010PhRvD..81f3008B}%
  \BibitemOpen
  \bibfield  {author} {\bibinfo {author} {\bibfnamefont {A.}~\bibnamefont
  {{B{\l}aut}}}, \bibinfo {author} {\bibfnamefont {S.}~\bibnamefont {{Babak}}},
  \ and\ \bibinfo {author} {\bibfnamefont {A.}~\bibnamefont {{Kr{\'o}lak}}},\
  }\href {\doibase 10.1103/PhysRevD.81.063008} {\bibfield  {journal} {\bibinfo
  {journal} {\prd}\ }\textbf {\bibinfo {volume} {81}},\ \bibinfo {eid} {063008}
  (\bibinfo {year} {2010})},\ \Eprint {http://arxiv.org/abs/0911.3020}
  {arXiv:0911.3020 [gr-qc]} \BibitemShut {NoStop}%
\bibitem [{\citenamefont {{Foreman-Mackey}}\ \emph {et~al.}(2013)\citenamefont
  {{Foreman-Mackey}}, \citenamefont {{Hogg}}, \citenamefont {{Lang}},\ and\
  \citenamefont {{Goodman}}}]{2013PASP..125..306F}%
  \BibitemOpen
  \bibfield  {author} {\bibinfo {author} {\bibfnamefont {D.}~\bibnamefont
  {{Foreman-Mackey}}}, \bibinfo {author} {\bibfnamefont {D.~W.}\ \bibnamefont
  {{Hogg}}}, \bibinfo {author} {\bibfnamefont {D.}~\bibnamefont {{Lang}}}, \
  and\ \bibinfo {author} {\bibfnamefont {J.}~\bibnamefont {{Goodman}}},\ }\href
  {\doibase 10.1086/670067} {\bibfield  {journal} {\bibinfo  {journal} {\pasp}\
  }\textbf {\bibinfo {volume} {125}},\ \bibinfo {pages} {306} (\bibinfo {year}
  {2013})},\ \Eprint {http://arxiv.org/abs/1202.3665} {arXiv:1202.3665
  [astro-ph.IM]} \BibitemShut {NoStop}%
\bibitem [{\citenamefont {{Speagle}}(2020)}]{2020MNRAS.493.3132S}%
  \BibitemOpen
  \bibfield  {author} {\bibinfo {author} {\bibfnamefont {J.~S.}\ \bibnamefont
  {{Speagle}}},\ }\href {\doibase 10.1093/mnras/staa278} {\bibfield  {journal}
  {\bibinfo  {journal} {\mnras}\ }\textbf {\bibinfo {volume} {493}},\ \bibinfo
  {pages} {3132} (\bibinfo {year} {2020})},\ \Eprint
  {http://arxiv.org/abs/1904.02180} {arXiv:1904.02180 [astro-ph.IM]}
  \BibitemShut {NoStop}%
\bibitem [{\citenamefont {Kass}\ and\ \citenamefont
  {Raftery}(1995)}]{Kass1995BayesFA}%
  \BibitemOpen
  \bibfield  {author} {\bibinfo {author} {\bibfnamefont {R.~E.}\ \bibnamefont
  {Kass}}\ and\ \bibinfo {author} {\bibfnamefont {A.~E.}\ \bibnamefont
  {Raftery}},\ }\href {\doibase 10.1080/01621459.1995.10476572} {\bibfield
  {journal} {\bibinfo  {journal} {Journal of the American Statistical
  Association}\ }\textbf {\bibinfo {volume} {90}},\ \bibinfo {pages} {773}
  (\bibinfo {year} {1995})}\BibitemShut {NoStop}%
\bibitem [{\citenamefont {{Normandin}}\ \emph {et~al.}(2018)\citenamefont
  {{Normandin}}, \citenamefont {{Mohanty}},\ and\ \citenamefont
  {{Weerathunga}}}]{2018PhRvD..98d4029N}%
  \BibitemOpen
  \bibfield  {author} {\bibinfo {author} {\bibfnamefont {M.~E.}\ \bibnamefont
  {{Normandin}}}, \bibinfo {author} {\bibfnamefont {S.~D.}\ \bibnamefont
  {{Mohanty}}}, \ and\ \bibinfo {author} {\bibfnamefont {T.~S.}\ \bibnamefont
  {{Weerathunga}}},\ }\href {\doibase 10.1103/PhysRevD.98.044029} {\bibfield
  {journal} {\bibinfo  {journal} {\prd}\ }\textbf {\bibinfo {volume} {98}},\
  \bibinfo {eid} {044029} (\bibinfo {year} {2018})},\ \Eprint
  {http://arxiv.org/abs/1806.01881} {arXiv:1806.01881 [astro-ph.IM]}
  \BibitemShut {NoStop}%
\bibitem [{\citenamefont {{Littenberg}}\ \emph {et~al.}(2020)\citenamefont
  {{Littenberg}}, \citenamefont {{Cornish}}, \citenamefont {{Lackeos}},\ and\
  \citenamefont {{Robson}}}]{2020PhRvD.101l3021L}%
  \BibitemOpen
  \bibfield  {author} {\bibinfo {author} {\bibfnamefont {T.~B.}\ \bibnamefont
  {{Littenberg}}}, \bibinfo {author} {\bibfnamefont {N.~J.}\ \bibnamefont
  {{Cornish}}}, \bibinfo {author} {\bibfnamefont {K.}~\bibnamefont
  {{Lackeos}}}, \ and\ \bibinfo {author} {\bibfnamefont {T.}~\bibnamefont
  {{Robson}}},\ }\href {\doibase 10.1103/PhysRevD.101.123021} {\bibfield
  {journal} {\bibinfo  {journal} {\prd}\ }\textbf {\bibinfo {volume} {101}},\
  \bibinfo {eid} {123021} (\bibinfo {year} {2020})},\ \Eprint
  {http://arxiv.org/abs/2004.08464} {arXiv:2004.08464 [gr-qc]} \BibitemShut
  {NoStop}%
\bibitem [{\citenamefont {{Karnesis}}\ \emph {et~al.}(2021)\citenamefont
  {{Karnesis}}, \citenamefont {{Babak}}, \citenamefont {{Pieroni}},
  \citenamefont {{Cornish}},\ and\ \citenamefont
  {{Littenberg}}}]{2021PhRvD.104d3019K}%
  \BibitemOpen
  \bibfield  {author} {\bibinfo {author} {\bibfnamefont {N.}~\bibnamefont
  {{Karnesis}}}, \bibinfo {author} {\bibfnamefont {S.}~\bibnamefont {{Babak}}},
  \bibinfo {author} {\bibfnamefont {M.}~\bibnamefont {{Pieroni}}}, \bibinfo
  {author} {\bibfnamefont {N.}~\bibnamefont {{Cornish}}}, \ and\ \bibinfo
  {author} {\bibfnamefont {T.}~\bibnamefont {{Littenberg}}},\ }\href {\doibase
  10.1103/PhysRevD.104.043019} {\bibfield  {journal} {\bibinfo  {journal}
  {\prd}\ }\textbf {\bibinfo {volume} {104}},\ \bibinfo {eid} {043019}
  (\bibinfo {year} {2021})},\ \Eprint {http://arxiv.org/abs/2103.14598}
  {arXiv:2103.14598 [astro-ph.IM]} \BibitemShut {NoStop}%
\bibitem [{\citenamefont {{Zhang}}\ \emph {et~al.}(2022)\citenamefont
  {{Zhang}}, \citenamefont {{Zhao}}, \citenamefont {{Mohanty}},\ and\
  \citenamefont {{Liu}}}]{2022PhRvD.106j2004Z}%
  \BibitemOpen
  \bibfield  {author} {\bibinfo {author} {\bibfnamefont {X.-H.}\ \bibnamefont
  {{Zhang}}}, \bibinfo {author} {\bibfnamefont {S.-D.}\ \bibnamefont {{Zhao}}},
  \bibinfo {author} {\bibfnamefont {S.~D.}\ \bibnamefont {{Mohanty}}}, \ and\
  \bibinfo {author} {\bibfnamefont {Y.-X.}\ \bibnamefont {{Liu}}},\ }\href
  {\doibase 10.1103/PhysRevD.106.102004} {\bibfield  {journal} {\bibinfo
  {journal} {\prd}\ }\textbf {\bibinfo {volume} {106}},\ \bibinfo {eid}
  {102004} (\bibinfo {year} {2022})},\ \Eprint
  {http://arxiv.org/abs/2206.12083} {arXiv:2206.12083 [gr-qc]} \BibitemShut
  {NoStop}%
\bibitem [{\citenamefont {{Lu}}\ \emph {et~al.}()\citenamefont {{Lu}},
  \citenamefont {{Li}}, \citenamefont {{Hu}}, \citenamefont {{Zhang}},\ and\
  \citenamefont {{Mei}}}]{2022arXiv220502384L}%
  \BibitemOpen
  \bibfield  {author} {\bibinfo {author} {\bibfnamefont {Y.}~\bibnamefont
  {{Lu}}}, \bibinfo {author} {\bibfnamefont {E.-K.}\ \bibnamefont {{Li}}},
  \bibinfo {author} {\bibfnamefont {Y.-M.}\ \bibnamefont {{Hu}}}, \bibinfo
  {author} {\bibfnamefont {J.-d.}\ \bibnamefont {{Zhang}}}, \ and\ \bibinfo
  {author} {\bibfnamefont {J.}~\bibnamefont {{Mei}}},\ }\href@noop {} {\
  }\Eprint {http://arxiv.org/abs/2205.02384 (2022)} {arXiv:2205.02384 (2022)
  [astro-ph.GA]} \BibitemShut {NoStop}%
\bibitem [{\citenamefont {{Cornish}}\ and\ \citenamefont
  {{Robson}}(2017)}]{2017JPhCS.840a2024C}%
  \BibitemOpen
  \bibfield  {author} {\bibinfo {author} {\bibfnamefont {N.}~\bibnamefont
  {{Cornish}}}\ and\ \bibinfo {author} {\bibfnamefont {T.}~\bibnamefont
  {{Robson}}},\ }in\ \href {\doibase 10.1088/1742-6596/840/1/012024} {\emph
  {\bibinfo {booktitle} {Journal of Physics Conference Series}}},\ \bibinfo
  {series} {Journal of Physics Conference Series}, Vol.\ \bibinfo {volume}
  {840}\ (\bibinfo {year} {2017})\ p.\ \bibinfo {pages} {012024},\ \Eprint
  {http://arxiv.org/abs/1703.09858} {arXiv:1703.09858 [astro-ph.IM]}
  \BibitemShut {NoStop}%
\bibitem [{\citenamefont {{Digman}}\ and\ \citenamefont
  {{Cornish}}(2022)}]{2022ApJ...940...10D}%
  \BibitemOpen
  \bibfield  {author} {\bibinfo {author} {\bibfnamefont {M.~C.}\ \bibnamefont
  {{Digman}}}\ and\ \bibinfo {author} {\bibfnamefont {N.~J.}\ \bibnamefont
  {{Cornish}}},\ }\href {\doibase 10.3847/1538-4357/ac9139} {\bibfield
  {journal} {\bibinfo  {journal} {\apj}\ }\textbf {\bibinfo {volume} {940}},\
  \bibinfo {eid} {10} (\bibinfo {year} {2022})},\ \Eprint
  {http://arxiv.org/abs/2206.14813} {arXiv:2206.14813 [astro-ph.IM]}
  \BibitemShut {NoStop}%
\bibitem [{\citenamefont {{Pieroni}}\ and\ \citenamefont
  {{Barausse}}(2020)}]{2020JCAP...07..021P}%
  \BibitemOpen
  \bibfield  {author} {\bibinfo {author} {\bibfnamefont {M.}~\bibnamefont
  {{Pieroni}}}\ and\ \bibinfo {author} {\bibfnamefont {E.}~\bibnamefont
  {{Barausse}}},\ }\href {\doibase 10.1088/1475-7516/2020/07/021} {\bibfield
  {journal} {\bibinfo  {journal} {\jcap}\ }\textbf {\bibinfo {volume} {2020}},\
  \bibinfo {eid} {021} (\bibinfo {year} {2020})},\ \Eprint
  {http://arxiv.org/abs/2004.01135} {arXiv:2004.01135 [astro-ph.CO]}
  \BibitemShut {NoStop}%
\bibitem [{\citenamefont {{Wang}}\ \emph {et~al.}(2022)\citenamefont {{Wang}},
  \citenamefont {{Li}}, \citenamefont {{Xu}},\ and\ \citenamefont
  {{Fan}}}]{2022PhRvD.106d4054W}%
  \BibitemOpen
  \bibfield  {author} {\bibinfo {author} {\bibfnamefont {G.}~\bibnamefont
  {{Wang}}}, \bibinfo {author} {\bibfnamefont {B.}~\bibnamefont {{Li}}},
  \bibinfo {author} {\bibfnamefont {P.}~\bibnamefont {{Xu}}}, \ and\ \bibinfo
  {author} {\bibfnamefont {X.}~\bibnamefont {{Fan}}},\ }\href {\doibase
  10.1103/PhysRevD.106.044054} {\bibfield  {journal} {\bibinfo  {journal}
  {\prd}\ }\textbf {\bibinfo {volume} {106}},\ \bibinfo {eid} {044054}
  (\bibinfo {year} {2022})},\ \Eprint {http://arxiv.org/abs/2201.10902}
  {arXiv:2201.10902 [gr-qc]} \BibitemShut {NoStop}%
\bibitem [{\citenamefont {{Boileau}}\ \emph {et~al.}()\citenamefont
  {{Boileau}}, \citenamefont {{Christensen}}, \citenamefont {{Gowling}},
  \citenamefont {{Hindmarsh}},\ and\ \citenamefont
  {{Meyer}}}]{2022arXiv220913277B}%
  \BibitemOpen
  \bibfield  {author} {\bibinfo {author} {\bibfnamefont {G.}~\bibnamefont
  {{Boileau}}}, \bibinfo {author} {\bibfnamefont {N.}~\bibnamefont
  {{Christensen}}}, \bibinfo {author} {\bibfnamefont {C.}~\bibnamefont
  {{Gowling}}}, \bibinfo {author} {\bibfnamefont {M.}~\bibnamefont
  {{Hindmarsh}}}, \ and\ \bibinfo {author} {\bibfnamefont {R.}~\bibnamefont
  {{Meyer}}},\ }\href@noop {} {\ }\Eprint {http://arxiv.org/abs/2209.13277
  (2022)} {arXiv:2209.13277 (2022) [gr-qc]} \BibitemShut {NoStop}%
\bibitem [{\citenamefont {{Cheng}}\ \emph {et~al.}()\citenamefont {{Cheng}},
  \citenamefont {{Li}}, \citenamefont {{Hu}}, \citenamefont {{Liang}},
  \citenamefont {{Zhang}},\ and\ \citenamefont {{Mei}}}]{2022arXiv220811615C}%
  \BibitemOpen
  \bibfield  {author} {\bibinfo {author} {\bibfnamefont {J.}~\bibnamefont
  {{Cheng}}}, \bibinfo {author} {\bibfnamefont {E.-K.}\ \bibnamefont {{Li}}},
  \bibinfo {author} {\bibfnamefont {Y.-M.}\ \bibnamefont {{Hu}}}, \bibinfo
  {author} {\bibfnamefont {Z.-C.}\ \bibnamefont {{Liang}}}, \bibinfo {author}
  {\bibfnamefont {J.-d.}\ \bibnamefont {{Zhang}}}, \ and\ \bibinfo {author}
  {\bibfnamefont {J.}~\bibnamefont {{Mei}}},\ }\href@noop {} {\ }\Eprint
  {http://arxiv.org/abs/2208.11615 (2022)} {arXiv:2208.11615 (2022) [gr-qc]}
  \BibitemShut {NoStop}%
\bibitem [{\citenamefont {{Boileau}}\ \emph {et~al.}(2022)\citenamefont
  {{Boileau}}, \citenamefont {{Jenkins}}, \citenamefont {{Sakellariadou}},
  \citenamefont {{Meyer}},\ and\ \citenamefont
  {{Christensen}}}]{2022PhRvD.105b3510B}%
  \BibitemOpen
  \bibfield  {author} {\bibinfo {author} {\bibfnamefont {G.}~\bibnamefont
  {{Boileau}}}, \bibinfo {author} {\bibfnamefont {A.~C.}\ \bibnamefont
  {{Jenkins}}}, \bibinfo {author} {\bibfnamefont {M.}~\bibnamefont
  {{Sakellariadou}}}, \bibinfo {author} {\bibfnamefont {R.}~\bibnamefont
  {{Meyer}}}, \ and\ \bibinfo {author} {\bibfnamefont {N.}~\bibnamefont
  {{Christensen}}},\ }\href {\doibase 10.1103/PhysRevD.105.023510} {\bibfield
  {journal} {\bibinfo  {journal} {\prd}\ }\textbf {\bibinfo {volume} {105}},\
  \bibinfo {eid} {023510} (\bibinfo {year} {2022})},\ \Eprint
  {http://arxiv.org/abs/2109.06552} {arXiv:2109.06552 [gr-qc]} \BibitemShut
  {NoStop}%
\end{thebibliography}%

\end{document}